\documentclass[journal]{IEEEtran}
\usepackage{graphicx,amssymb,amsmath}
\usepackage[noadjust]{cite}
\usepackage{setspace}               
\usepackage{stfloats}
\usepackage{caption}
\usepackage{subfigure}

\usepackage{graphics}
\usepackage{enumerate}

\graphicspath{ {./images/} }
\usepackage{amsthm}
\usepackage{flushend}
\usepackage{amsmath}
\usepackage{cases,subeqnarray}
\usepackage{bm,bigstrut}
\usepackage{textcomp}
\usepackage{latexsym,bm}
\usepackage{booktabs,changebar}
\usepackage{xcolor}
\usepackage{mathtools}
\usepackage{dsfont}
\usepackage{extarrows}
\usepackage{mathrsfs}
\usepackage{cleveref}
\usepackage{url}
\usepackage{multicol}       
\usepackage{multirow}       
\usepackage{array}          
\definecolor{crimson}{RGB}{192,0,0}         
\definecolor{navy}{RGB}{47,85,151}         
\usepackage{geometry}
\makeatletter                               
\newif\if@restonecol
\makeatother

\makeatletter
\newif\if@restonecol
\makeatother

\usepackage[linesnumbered,ruled,vlined]{algorithm2e}
\usepackage{algpseudocode}
\usepackage{amsmath}
\theoremstyle{plain}

\theoremstyle{plain}

\begin{document}

	\title{Low-Complexity Precoding for Extremely Large-Scale MIMO Over Non-Stationary Channels}
	\author{Bokai Xu, Zhe Wang, Huahua Xiao, Jiayi Zhang,~\IEEEmembership{Senior Member,~IEEE,}\\
	 Bo Ai,~\IEEEmembership{Fellow,~IEEE},
	and Derrick Wing Kwan Ng,~\IEEEmembership{Fellow,~IEEE}
		
			\thanks{B. Xu, Z. Wang, and J. Zhang are with the School of Electronic and Information Engineering, Beijing Jiaotong University, Beijing 100044, China. (e-mail: \{20251197, zhewang\_77, jiayizhang\}@bjtu.edu.cn).}
			\thanks{H. Xiao is with ZTE Corporation, State Key Laboratory of Mobile Network and Mobile Multimedia Technology.}
           \thanks{B. Ai is with the State Key Laboratory of Rail Traffic Control and Safety, Beijing Jiaotong University, Beijing 100044, China.}
            \thanks{D. W. K. Ng is with School of Electrical Engineering and Telecommunications, University of New South Wales, NSW 2052, Australia.}
	}
	\maketitle
	
	\begin{abstract}
		Extremely large-scale multiple-input-multiple-output (XL-MIMO) is a promising technology for the future sixth-generation (6G) networks to achieve higher performance. In practice, various linear precoding schemes, such as zero-forcing (ZF) and regularized zero-forcing (RZF) precoding, are capable of achieving both large spectral efficiency (SE) and low bit error rate (BER) in traditional massive MIMO (mMIMO) systems. However, these methods are not efficient in extremely large-scale regimes due to the inherent spatial non-stationarity and high computational complexity. To address this problem, we investigate a low-complexity precoding algorithm, e.g., randomized Kaczmarz (rKA), taking into account the spatial non-stationary properties in XL-MIMO systems. Furthermore, we propose a novel mode of randomization, i.e., sampling without replacement rKA (SwoR-rKA), which enjoys a faster convergence speed than the rKA algorithm. Besides, the closed-form expression of SE considering the interference between subarrays in downlink XL-MIMO systems is derived. Numerical results show that the complexity given by both rKA and SwoR-rKA algorithms has $51.3 \%$ reduction than the traditional RZF algorithm with similar SE performance. More importantly, our algorithms can effectively reduce the BER when the transmitter has imperfect channel estimation.
	\end{abstract}
	
	\section{Introduction}
	Advanced novel wireless technologies have been developed, e.g., reconfigurable intelligent surface (RIS) \cite{9113273} and cell-free massive MIMO (CF-MIMO) \cite{8768014,9174860} to satisfy the key performance indicators (KPI) set for the future sixth-generation (6G) networks \cite{cui2022near,9690475}. One of these key technologies is the extremely large-scale MIMO (XL-MIMO) by equipping an extremely large number of antennas that can dramatically improve spectral efficiency (SE). As such, numerous XL-MIMO hardware design schemes \cite{Wang}, e.g., extremely large antenna array (ELAA), large intelligent surfaces (LIS), etc., have been proposed to improve the spatial degrees of freedom (DoF) for achieving a higher area throughput \cite{emil,dardari2020communicating,9733790,9416909,9453784,9743355,9849114,zheng2022asynchronous,9737367,8995606}. In practice, with the number of antennas increasing, various new features of the channel emerge, e.g., the spatial non-stationary property. In particular, with a long antenna array, different parts of it may experience different characteristics. To exploit the effect of spatial non-stationary properties of channels, the BS antenna array can be divided into several subarrays. In particular, users only view the portion of the array, which is called Visibility Regions (VR), for addressing the discussed spatial channel variations \cite{amiri2019message,9718019}.
	On the other hand, the application of large number of antennas would also lead to higher computational complexity costs. Therefore, low-complexity precoding algorithms are vital for the successful implementation of practical XL-MIMO systems.
	
	 Most of the prior works were based on the statistical characteristics of the channels which may not be available in XL-MIMO systems. Besides, they often neglected the spatial non-stationary property \cite {boroujerdi2018low}. Moreover, to reduce the complexity of the signal processing,  some new algorithms, e.g., variational message passing (VMP) \cite{amiri2019message,9551696}, mean-angle-based zero-forcing (MZF), and tensor ZF (TZF) \cite{9616090} have been proposed. To the best knowledge of the authors, even though the above algorithms based on the ZF have low signal processing complexity, the SE performance is still unsatisfactory.  So it is necessary to design an algorithm to offer a possibility for a performance-complexity trade-off. On the other hand, the work in \cite{9145378, croisfelt2021accelerated} based on the regularized ZF (RZF) algorithm design achieves a better SE performance than the ZF algorithm. However, it focused on the low-complexity algorithm of uplink signal detection. Moreover, the interference between the subarrays is not considered.

	Motivated by the above observations, we aim at comparing the randomized Kaczmarz algorithm (rKA) and the sampling without replacement rKA (SwoR-rKA) with the conventional RZF algorithm and its goal is to find the most competitive precoding scheme.

	 $\bullet$ Firstly, we apply the rKA iteration algorithm to reduce the complexity of the traditional RZF matrix inversion in downlink XL-MIMO systems considering the spatial non-stationary property.

	  $\bullet$ Secondly, we propose an optimized scheme of the rKA, the SwoR-rKA algorithm, which has a faster convergence speed than the rKA algorithm. If the complexity is a concern, then the rKA algorithm can be used since the rKA has the lowest complexity. The price to pay is that the convergence speed gets a slight reduction.
	  
	   $\bullet$ Finally, we analyze the SE performance and bit error rate (BER) that consider the interference between subarrays. It is worth noting that the SwoR-rKA algorithm has the highest SE performance. Furthermore, both the rKA and the SwoR-rKA algorithms can reduce the BER when the transmitter has an imperfect channel estimate.

	{\emph {Organization:}} The rest of this paper is organized
	as follows. In Section \ref{sec:basic}, the XL-MIMO system model and non-stationary channel are introduced. In Section \ref{sec:algorithm}, we provide the details of designing the rKA precoding and the SwoR-rKA precoding. Numerical results\footnote{Simulation codes are provided to reproduce the results in this paper: https://github.com/BJTU-MIMO}
 are carried out in Section \ref{sec:simulation} and useful conclusions are drawn in Section \ref{sec:conclusion}.

	{\emph {Notation:}} Lowercase letters in bold, $\bf x$, denote column
	vectors and capital letters in bold, $\bf X$, denote matrices.
	$(\cdot )^{\text {H}}$ represents conjugate transpose. ${\bf I}_{n}$ is the $n \times n$ identity matrix. $\mathbb{E}\left \{ \cdot  \right \} $, $\text{diag}\left \{ \cdot  \right \}$, and $\text{tr}\left \{ \cdot  \right \} $ are the expectation operator, diagonalization operator, and trace operator, respectively. $\text{Var}(\cdot )$ denotes the operation of variance. Concatenating ${\bf Y}$ underneath matrix $\bf X$ is represented by $[{\bf X};{\bf Y}]$. $\left \| {\bf X} \right \|_{F}  =\sqrt{\left \langle {\bf X},{\bf X} \right \rangle } $ and $\left \| {\bf x} \right \| $ represents the Frobenius norm of the matrix $\bf X$ and the $l_{2}$-norm of the vector $\bf x$, respectively. $\mathcal{CN}( \mu ,\sigma)$ denotes the Gaussian distribution with mean $\mu$ and covariance $\sigma$.
	
		\section{System Model}\label{sec:basic}
	
\begin{figure}[t]
		\centering
		\includegraphics[width=2.5 in]{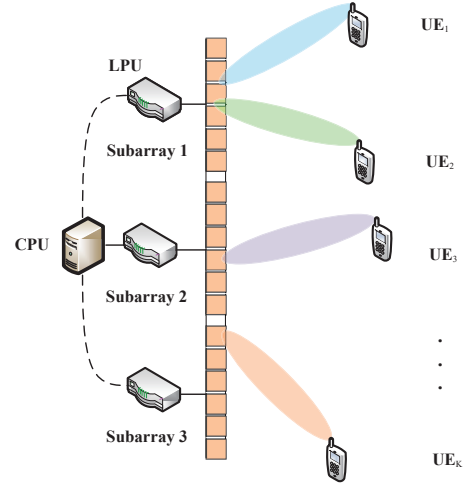}
		
		\caption{Illustration of the investigated XL-MIMO system.
		\label{fig:xl mimo}}
	\end{figure}

	As illustrated in Fig. \ref{fig:xl mimo}, we investigate the downlink of XL-MIMO systems, where a BS is equipped with $M$ antennas, serving $K$ single-antenna user equipments (UEs). Note that the $M$ antennas for the BS are divided into disjoint parts such that each part is with $M^{(s)}=M/S $ antennas, where $ {\textstyle \sum_{s=1}^{S}}M^{(s)}=M $. It is worth mentioning that the subarrays are equipped with their respective local processing units (LPUs) for conducting the related processing and all these LPUs are connected to a central processing unit (CPU). Besides, the $K$ UEs are evenly distributed in the $S$ subarrays, where $ {\textstyle \sum_{s=1}^{S}}K^{(s)}=K $. The received complex signal $y_{jk}\in \mathbb{C}$ at the $k$-th UE in the $j$-th subarray is given by
	\begin{equation}
y_{jk} =\sum_{s=1 }^{S} ({\bf h}_{jk}^{s})^{{\text H}}{\bf x}_{s} +n_{jk}
	\end{equation}	
	where ${\bf x}_{s} \in \mathbb{C}^{M^{(s)} \times 1} $ is the transmit signal in the $s$-th subarray and ${{\bf h}_{jk}^{s} }\in \mathbb{C}^{M^{(s)} \times 1 } $ represents the channel vector between the BS in the $s$-th subarray and the $k$-th UE in the $j$-th subarray, respectively. The additive circularly symmetric complex Gaussian noise at the $k$-th UE in the $j$-th subarray is denoted by $n_{jk}  \sim \mathcal{CN}(0,\sigma ^{2})$, $\forall k=1,\cdot \cdot \cdot ,K^{(j)}$, $\forall j=1,\cdot \cdot \cdot ,S$, where $\sigma ^{2}$ is the noise variance at the receiver.
	\vspace{-0.5em}
	Note that the BS can implement Gaussian codebooks and the linear precoding scheme such that the transmit signal ${\bf x}_{s}$ in (1) can be denoted as
	\begin{equation}
		{\bf x}_{s}=\sum_{i=1}^{K^{(s)}} {\bf g}_{si} s_{si}  ={\bf G}_{s} {\bf s}_{s},
	\end{equation}
	where ${\bf G}_{s}=[{\bf g}_{s1},\cdot \cdot \cdot ,{\bf g}_{sK^{(s)}}]\in \mathbb{C}^{M^{(s)}\times K^{(s)}} $ is the precoding matrix for $K^{(s)}$ UEs in the $s$-th subarray, $ {\bf g}_{si} $ $\in \mathbb{C}^{M^{(s)}\times 1}  $ denotes the precoding vector for the $i$-th UE in the $s$-th subarray and ${\bf s}_{s}=[s_{s1},\cdot \cdot \cdot ,s_{sK^{(s)}} ]^{\text{T}}$ is the complex vector containing the data symbols for all UEs in the $s$-th subarray.
	Consequently, the received signal in (1) can be expressed as
	\begin{equation}
	\begin{aligned}
y_{jk} &=\underbrace{({\bf h}^{j}_{jk})^{\text {H}} {\bf g}_{jk}s_{jk}}_{\text{Desired signal}}+\underbrace{\sum_{\mathop{i=1}\limits_{i\ne k} }^{K^{(j)}}({\bf h}^{j}_{jk})^{\text {H}} {\bf g}_{ji}s_{ji}}_{\text{Intra-subarray interference}}\\
		&+\underbrace{\sum_{\mathop{s=1}\limits_{s\ne j} }^{S}\sum_{i=1 }^{K^{(s)}}({\bf h}^{s}_{jk})^{\text {H}} {\bf g}_{si}s_{si}}_{\text{Inter-subarray interference}}+  \underbrace{n_{jk}}_{\text{Noise}},
	\end{aligned}
	\end{equation}
	where $s_{jk}\sim \mathcal{CN}(0,p_{jk} )  $ and ${\bf g}_{jk}$ satisfies the power constraint $\mathbb{E} \{ \left \| {\bf g}_{jk} \right \|^{2} \}=1$.
	We denote by 
 \begin{equation}
 {\bf H}^{s}_{j} = [{\bf h}_{j1}^{s},\cdot \cdot \cdot ,{\bf h}_{jK^{(j)}}^{s}]\in \mathbb{C}^{M^{(s)}\times K^{(j)}}
 \end{equation}
 the channel matrix between the $M^{(s)}$ antennas in the $s$-th subarray and $K^{(j)}$ UEs in the $j$-th subarray.

	\subsection{Non-Stationary Channel Model}
	In this paper, we consider a channel model with the non-stationary property. The channel vector between UE $k$ in the $j$-th subarray and $M^{(s)}$ antennas in the $s$-th subarray is denoted by \cite{ali2019linear}
	\begin{equation}
		{\bf h}_{jk}^{s} =\sqrt{M^{(s)}}({\bf \Theta} _{jk}^{s}) ^{\frac{1}{2} }{\bf z}_{jk}^{s},
	\end{equation}
	where ${\bf h}_{jk}^{s}\in  \mathbb{C}^{ M^{(s)}\times 1} $, ${\bf z}_{jk}^{s}\sim \mathcal{CN}({\bf 0},\frac{1}{M^{(s)}} {\bf I}_{M^{(s)}} ) \in  \mathbb{C}^{ M^{(s)} \times 1}$, and 
 \begin{equation}
 {\bf \Theta} _{jk}^{s} =({\bf D}_{jk}^{s})^{\frac{1}{2} }{\bf R}_{jk}^{s}({\bf D}_{jk}^{s})^{\frac{1}{2}} \in \mathbb{C}^{ M^{(s)} \times M^{(s)}},
 \end{equation}
 where ${\bf R}_{jk}^{s}\in \mathbb{C}^{M^{(s)}\times M^{(s)}}$ is the spatial correlation matrix between UE $k$ in the $j$-th subarray and BS in the $s$-th subarray. ${\bf D}_{jk}^{s}\in \mathbb{C}^{M^{(s)}\times M^{(s)}}$ is a diagonal matrix and has $D_{jk}^{s}$ non-zero diagonal elements between UE $k$ in the $j$-th subarray and BS in the $s$-th subarray.

	We consider two types of channel normalization schemes in non-stationary channels  \cite{ali2019linear}. In normalization 1, we assume that non-stationary channels have the same norm as stationary channels, which satisfies $\text{tr} \{{\bf \Theta}_{jk}^{s} \}  =\text{tr}\{{\bf R}_{jk}^{s} \} =M^{(s)} $ and $D_{jk}^{s}=\text{diag} \{[{\bf 0},\sqrt{\frac{M^{(s)} }{D_{jk}^{s} } }{\bf 1}_{D_{jk}^{s} },{\bf 0}]^{\text {H}}   \} $ in the $s$-th subarray. In normalization 2, the energy of the UE served by different numbers of antennas may be the same. Non-stationary channels have norm less than or equal to stationary ones. We assume that $\text{tr} \{{\bf \Theta}_{jk}^{s} \}=D_{jk}^{s}$ in the $s$-th subarray, where $D_{jk}^{s}=\text{diag} \{[{\bf 0},{\bf 1}_{D_{jk}^{s} },{\bf 0}]^{\text{H}} \}$.
	\subsection{CSI Assumption}
	We assume that the BS has an imperfect estimate of the channel for UE $k$ in the $s$-th subarray as
	\begin{equation}
		\hat{{\bf h}}_{jk}^{s}=({\bf \Phi}_{jk}^{s}) ^{\frac{1}{2} }( \sqrt{1-\tau ^{2}}{\bf z}_{jk}^{s}+\tau {\bf v}_{jk}^{s}  )= \sqrt{1-\tau ^{2}}{\bf h}_{jk}^{s}+\tau {\bf n}_{jk}^{s},
	\end{equation}
	where ${\bf h}_{jk}^{s}$ is the actual channel, ${\bf v}_{jk}^{s} \sim \mathcal{CN}({\bf 0}_{M^{(s)}\times 1} ,{\bf I}_{M^{(s)}} )$, and ${\bf n}_{jk}^{s} =({\bf \Phi}_{jk}^{s}) ^{\frac{1}{2} } {\bf v}_{jk}^{s}\sim \mathcal{CN}({\bf 0}_{M^{(s)}\times 1} ,{\bf \Phi}_{jk}^{s} ) $ is the independent error, respectively.
	The channel matrix between the BS in the $s$ subarray and $K^{(j)}$ users in the $j$ subarray is $  \hat {\bf H}_{j}^{s} = [ \hat {\bf h}_{j1}^{s},\cdot \cdot \cdot,\hat {\bf  h}_{jK^{(j)}}^{s}]\in \mathbb{C}^{M^{(s)}\times K^{(j)}}$.

	\begin{algorithm}[t]
		\caption{Low-Complexity Precoding Scheme Designed by the rKA Algorithm}
		\label{algo:review}
		\KwIn{The number of UEs $K^{(j)}$ in the $j$ subarray, the number of subarray antennas $M^{(j)}$, the inverse of the SNR $\xi \ge 0 $, the subarray channel matrix ${\bf H}_{j}^{j}\in \mathbb{C}^{M^{(j)}\times K^{(j)} }$, and the number of the algorithm iterations $T$;}
		{\bf Initiation:} ${\bf W}_{j}\in  \mathbb{C}^{K^{(j)}\times K^{(j)}}={\bf 0}$;\\
		\For{$k=1,2,...,K^{(j)}$}
		{
			Define the state vectors ${\bf m}_{j}^{t} \in \mathbb{C}^{M^{(j)} }$ and ${\bf n}_{j}^{t}\in \mathbb{C}^{K^{(j)}} $ with ${\bf m}_{j}^{0} ={\bf 0} $ and ${\bf n}_{j}^{0} ={\bf 0} $;\\
			Define user canonical basis ${\bf e}_{k}\in \mathbb{R}^{K^{(j)}}$, where $[{\bf e}_{k}]_{k}=1$ and $[{\bf e}_{k}]_{j}=0,\forall j\ne k $;\\
			\For{$t=0,1,...,T-1$}
			{

				Select the $r(t)$-th row of $({\bf H}_{j}^{j})^{\text {H}}$ with $r(t)\in\left \{1,2,...,K^{(j)} \right \}$ which satisfy $P_{r(t)}^{j}=1/K^{(j)}$, where $K^{(j)}$ denotes the number of users in the $j$-th subarray;

				Compute the residual:
				$ \eta ^{t}:=\frac{[{\bf e}_{k}]_{r(t)}-\left \langle {\bf h}_{jr(t)}^{j} ,{\bf m}_{j}^{t}  \right \rangle-\xi  n_{jr(t)}^{t}    }{\left \| {\bf h}_{jr(t)}^{j}  \right \|_{2}^{2}+\xi   }$;\\
				Update ${\bf m}_{j}^{t+1}={\bf m}_{j}^{t} +\eta ^{t}{\bf h}_{jr(t)}^{j} $;\\
				Update $n_{jr(t)}^{t+1}= n_{jr(t)}^{t}+\eta ^{t} $;\\
				Repeat $ n_{jr(t)}^{t+1} =n_{jr(t)}^{t} , \forall j\ne r(t)$;
			}
			Update $[{\bf W}^{(s)}]_{:,k}={\bf n}^{T-1 }$;
		}
		
		\KwOut{${\bf G }_{j}^{\text{RZF}}=\beta {\bf H }^{j}_{j}{\bf W}_{j}$; }
		
	\end{algorithm}

	\section{Randomized Kaczmarz Signal Precoding}\label{sec:algorithm}
	In this section, we consider two signal precoding schemes, the rKA algorithm and an improved version of the rKA, SwoR-rKA. Moreover, we derive the expression of the SE.
	\subsection{rKA Algorithm}
	The rKA method \cite{strohmer2006randomized} for solving a linear system of equations
	${\bf A}{\bf x}={\bf b}$ has been widely adopted.
	We first consider linear precoding schemes such as ZF and RZF \cite{9622183,9529197}. If the channel conditions are good, the precoding matrix ${\bf G}_{j}^{\text {RZF}}$ in the $j$-th subarray can be computed according to
	\begin{equation}
		{\bf G}_{j}^{\text {RZF}}=\beta {\bf H}^{j}_{j}(({\bf H}^{j}_{j})^{\text {H}}{\bf H}^{j}_{j}+\xi {\bf I}_{K^{(j)}}  )^{-1},
	\end{equation}
	where $\xi=\frac{1}{\text{SNR}}=\frac{\sigma ^{2} }{P} $ is the regularization factor. We define 
	\begin{equation}
	{\bf F}_{j}={\bf H}^{j}_{j}(({\bf H}^{j}_{j})^{\text {H}}{\bf H}^{j}_{j}+\xi {\bf I}_{K^{(j)}}  )^{-1}
	\end{equation}
	 and 
  \begin{equation}
  \beta=\sqrt{P / \operatorname{Tr}\left({\bf F}_{j}^{H} {\bf F}_{j}\right)} >0
  \end{equation}
  represents the power control factor.
	Then, the BS in the $j$-th subarray transmits the $M^{(j)}$-dimension vector via its $M^{(j)}$ antennas, which is
	\begin{equation}
		{\bf x}_{j}={\bf G}_{j}{\bf s}_{j}=\beta {\bf H}^{j}_{j}(({\bf H}^{j}_{j})^{\text {H}}{\bf H}^{j}_{j}+\xi {\bf I}_{K^{(j)}}   )^{-1}{\bf s}_{j}. 	
	\end{equation}

	We can realize (8) with low computational complexity when the antenna size and subarray are small. However, for XL-MIMO systems, the computational cost of the matrix inversion can be exceedingly high. To address this issue, we apply the rKA algorithm based on iterative matrix factorization. We define $K^{(j)}$-dim signal 
	\begin{equation}
	{\bf w}_{j}=(({\bf H}^{j}_{j})^{\text {H}}{\bf H}^{j}_{j}+\xi {\bf I}_{K^{(j)}}   )^{-1}{\bf s}_{j}.
	\end{equation}
	Exploiting the rKA algorithm, we can transform the former expression in (12) into an optimization problem
	\begin{equation}
		{\bf w}_{j}=\underset{{\bf x}_{j}\in  \mathbb{C}^{K^{(j)}}}{\arg \min }\parallel {\bf H}^{j}_{j}{\bf x}_{j} \parallel^{2}+\xi  \parallel {\bf x}_{j}-{\bf s}_{j}^{\xi} \parallel^{2},
	\end{equation}
	where ${\bf s}_{j}^{\xi}=\frac{{\bf s} _{j}}{\xi}$ is the transmitted signal combining the regularization factor in the $j$-th subarray.
	
	\begin{IEEEproof}[Proof]
	 We set the derivation of (13) with respect to ${\bf x}_{j}$ ($\frac{\partial {\bf w}_{j}}{\partial {\bf x}_{j}}$) as $\bf 0$:
	\begin{equation}
		\frac{\partial {\bf w}_{j}}{\partial {\bf x}_{j}}={({\bf H}^{j}_{j} )}^{\text {H}}{\bf H}^{j}_{j} {\bf x}_{j}+\xi {\bf x}_{j}-{\bf s}_{j}={\bf 0}, 	
	\end{equation}
	so that we can obtain the optimal solution as 
	\begin{equation}
		{\bf x}^{*}_{j}=({({\bf H}^{j}_{j} )}^{\text {H}}{\bf H}^{j}_{j} +\xi {\bf I}_{K^{(j)}}  )^{-1}{\bf s}_{j}.
	\end{equation}
	So we have ${\bf w}_{j}=({({\bf H}^{j}_{j} )}^{\text {H}}{\bf H}^{j}_{j}+\xi {\bf I}_{K^{(j)}}  )^{-1}{\bf s}_{j}$.
\end{IEEEproof}
\renewcommand{\IEEEQED}{\IEEEQEDopen}
	Then (12) can be expressed as \vspace{-0.5em}
	\begin{equation}
		{\bf w}_{j}=\underset{{\bf x}_{j}\in  \mathbb{C}^{K^{(j)}}}{\arg \min }\parallel {\bf A}_{j}{\bf x}_{j}-{\bf a}_{j}\parallel ^{2},
	\end{equation}
	where ${\bf A}_{j}=[{\bf H}_{j}^{j} ; \sqrt{\xi}{\bf I}_{K^{(j)}}]\in \mathbb{C}^{(M^{(j)}+K^{(j)})\times K^{(j)} } $, ${\bf a}_{j}=[{\bf 0};\sqrt{\xi }{\bf s}_{j}^{\xi}]\in \mathbb{C}^{(M^{(j)}+K^{(j)})}$, and $\sqrt{\xi }{\bf s}_{j}^{\xi}=\frac{{\bf s}_{j} }{\sqrt{\xi } } $, respectively. The set of linear equations (SLE) ${\bf A}_{j}{\bf x}_{j}={\bf a}_{j}$ is over-determined (OD) and should be solved for the vector ${\bf x}_{j}\in \mathbb{C}^{K^{(j)}}  $. This SLE is inconsistent unless ${\bf s}_{j}={\bf 0}$,
	so the above SLE can be denoted by \vspace{-0.5em}
	\begin{equation}
		({\bf A}_{j})^{\text {H}}{\bf z}_{j}=({\bf A}_{j})^{\text {H}}{\bf a}_{j}=\sqrt{\xi }\frac{{\bf s}_{j}}{\sqrt{\xi } }={\bf s}_{j},
	\end{equation}
	where we define ${\bf z}_{j}={\bf A}_{j}{\bf w}_{j}$. Denoting by ${\bf z}_{j}^{t}=[{\bf m}_{j}^{t};\sqrt{\xi }{\bf n}_{j}^{t}]$ at each iteration $t$, then we can obtain ${\bf w}_{j}$ in two steps:
	\begin{enumerate}[\underline {\bf Step 1}:]
	\item We solve the SLE
     \begin{equation}
     	({\bf A}_{j})^{\text {H}}{\bf z}_{j}={\bf s}_{j}
     \end{equation}
	We execute $K^{(j)}$ rKA algorithm in parallel and input to the $k$-th rKA is ${\bf s}_{j}={\bf e}_{k}\in \mathbb{C}^{K^{(j)}} $, where ${\bf e}_{k}$ denotes the $k$-th canonical basis \cite{boroujerdi2018low}.
	It is noticed that we select the $r(t)$-th row of $({\bf H}_{j}^{j})^{\text {H}}$ with the possibility
	\begin{equation}
		P_{r(t)}^{j}=1/K^{(j)},
	\end{equation}	
	where $K^{(j)}$ is the number of UEs in the $j$-th subarray.
	
	\end{enumerate}

	\begin{enumerate}[\underline {\bf Step 2}:]
	\item We divide the last $K^{(j)}$ components of ${\bf z}_{j}$ by $\sqrt{\xi }$, i.e., ${\bf w}_{j}={\bf n}_{j}$.
	\end{enumerate}
	When we run $K^{(j)}$ (18) by rKA in parallel, we can obtain $K^{(j)}$ ${\bf w}_{jk}$, where $k=1, \cdot \cdot \cdot, K^{(j)}$. We define the matrix 
	\begin{equation}
	{\bf W}_{j}=[{\bf w}_{j1},\cdot \cdot \cdot,{\bf w}_{jK^{(j)}} ]\in \mathbb{C}^{K^{(j)}\times K^{(j)}},
	\end{equation}
	 so the RZF matrix for the downlink in (8) can be approximated as \vspace{-0.5em}
	\begin{equation}
		{\bf G }_{j}^{\text{RZF}}=\beta {\bf H }^{j}_{j}{\bf W}_{j}
	\end{equation}
	and the precoding vector in the $j$-th subarray is \vspace{-0.5em}
	\begin{equation}
		{\bf x}_{j}={\bf G}_{j}^{\text{RZF}}{\bf s}_{j}=\beta {\bf H}^{j}_{j}{\bf W}_{j}{\bf s}_{j}.
	\end{equation}
 We summarize the above steps in Algorithm \ref{algo:review}, and the convergence analysis of the Algorithm \ref{algo:review} is provided in Appendix \ref{secA}.
	\begin{table}[t]
		\centering
		\fontsize{9}{12}\selectfont
		\caption{Simulation Parameters in XL-MIMO Systems.}
		\label{tab:Simulation parameters}
		\begin{tabular}{|p{1.5cm}<{\centering}|p{1.5cm}<{\centering}|p{2.5cm}<{\centering}|p{1.8cm}<{\centering}|}
			\hline
			\bf Parameter & \bf Value & \bf Parameter & \bf Value\cr\hline
			\hline
			
			$M$ & 256 &Array type& ULA \cr \hline

			$K$ & 16 &Carrier frequency & 2.6 GHz \cr \hline

			$S$& 4 &SNR  & [5, 20] dBm \cr  \hline

			$M^{(s)} $& 64 &$\tau $ & $0.3$ \cr  \hline

			$\sigma ^{2} $& 1 dBm &$T$& $150$ \cr  \hline
						
		\end{tabular}
	\end{table}

	\subsection{SwoR-rKA Algorithm}
	To accelerate the speed of the convergence in Algorithm 1, some modified versions of the rKA-based schemes can be implemented \cite{eldar2011acceleration}.
	To achieve higher SE and accelerate the speed of the convergence, inspired by \cite{croisfelt2021accelerated}, we apply the SwoR-rKA algorithm with a different randomization scheme compared to the former rKA algorithm. Before the iteration from step 2 to step 11 in Algorithm \ref{algo:review}, we store $\left \|{\bf h}_{jr(t)}^{j}   \right \|_{2}^{2}+\xi   $ and $\left \| {\bf H}^{j}_{j}  \right \|_{F}^{2}+K^{(j)}\xi$. We replace (19) with
	\begin{equation}
		P_{r(t)}^{j}=\frac{\left \|{\bf h}_{jr(t)}^{j}   \right \|_{2}^{2}+\xi   }{\left \| {\bf H}^{j}_{j}  \right \|_{F}^{2}+K^{(j)} \xi},
		\end{equation}
	when the $P_{r(t)}^{j}$ is replaced by (23), we call the rKA algorithm in Algorithm \ref{algo:review} as ``SwoR-rKA'' algorithm.
	
	By analyzing the algorithm, we observe that the proposed SwoR-rKA can pick a different ${\bf h}_{jr(t)}^{j}$ and ${\bf n}_{jr(t)}^{t}$ each time, increasing the probability of selecting users which has better channel conditions compared with equal probability selection in rKA algorithm. Motivated by these advantages provided by the SwoR-rKA algorithm, we apply the SwoR-rKA algorithm to XL-MIMO systems.
	\vspace{-0.5em}
	\subsection{SE Analysis}
	\vspace{-0.5em}
	For a discrete memoryless interference channel, output $y\in \mathbb{C}$ is
	\begin{equation}
		y=hx+\gamma +n,
	\end{equation}
	where $n\sim  \mathcal{CN}(0,\sigma^{2}  )$ and $\gamma$ is the
	interference term. Then the channel capacity $C$ is
		\begin{equation}
		C\ge \mathbb{E}\left \{ \text {log}_{2}(1+\frac{p\left |h  \right |^{2}  }{p_{\gamma} +\sigma ^{2} } )  \right \},
	\end{equation}
	where $h=\mathbb{E} \{ ({\bf h}_{jk}^{j} )^{\text {H}}{\bf g}_{jk}   \}$. The signal-to-interference-and-noise ratio (SINR) is defined as $\frac{p\left | h \right |^{2}  }{p_{\gamma}+\sigma ^{2} }$. Then, we consider the interference term $\gamma$ according to (3) for the BS in the $j$-th subarray and UEs in the $s$-th subarray
	\begin{equation}
		\begin{aligned}
	\gamma&=\left( ({\bf h}_{jk}^{j} )^{\text {H}}{\bf g}_{jk}-\mathbb{E} \{ ({\bf h}_{jk}^{j} )^{\text {H}}{\bf g}_{jk}   \}\right ) s_{jk}\\ &+\sum_{\mathop{i=1}\limits_{ i\ne k}}^{K^{(j)}  }({\bf h}_{jk}^{j} )^{\text {H}}{\bf g}_{ji}s_{ji}+\sum_{\mathop{s=1}\limits_{ s\ne j}}^{S }\sum_{i=1}^{K^{(s)}}(h_{jk}^{s} )^{\text{H}}{\bf g}_{si}s_{si}\\ &= \sum_{s=1}^{S}\sum_{i=1}^{K^{(s)} }({\bf h}_{jk}^{s} )^{\text {H}}{\bf g}_{si}s_{si}-\mathbb{E}\{({\bf h}_{jk}^{s} )^{\text {H}}{\bf g}_{jk} \}s_{jk}.
	\end{aligned}
	\end{equation}
  Furthermore, the variance of the interference term is
	\begin{equation}
	\begin{aligned}
p_{\gamma}&=\mathbb{E}\left\{|\gamma|^{2}\right\} \\
&=\sum_{s=1}^{S} \sum_{i=1}^{K^{(s)}} \mathbb{E}\left\{|\left(\mathbf{h}_{j k}^{s}\right)^{\mathrm{H}} \mathbf{g}_{si}|^{2}\right\} \mathbb{E}\left\{\left|s_{s i}\right|^{2}\right\}\\ &-\left|\mathbb{E}\{(\mathbf{h}_{j k}^{j})^{\mathrm{H}} \mathbf{g}_{j k}\}\right|^{2} \mathbb{E}\left\{\left|s_{j k}\right|^{2}\right\} \\
&=\sum_{s=1}^{S} \sum_{i=1}^{K^{(s)}} p_{s i} \mathbb{E}\left\{\left|\mathbf{g}_{s i}^{\mathrm{H}}\mathbf{h}_{j k}^{s}\right|^{2}\right\}-p_{j k}\left|\mathbb{E}\{\mathbf{g}_{j k}^{\mathrm{H}} \mathbf{h}_{j k}^{j}\}\right|^{2}.
\end{aligned}
\end{equation}	
	So the SINR in the $j$-th subarray is
	\begin{small}
	\begin{equation}
		\text{SINR}_{jk}\!=\frac{\overbrace{p_{jk}\!\mid \mathbb{E} \{{\bf g}_{jk}^{\text {H}}{\bf h}^{j}_{jk}\}\! \mid ^{2}}^{\text{Received signal power}}}{\underbrace{ \!\sum_{\mathop{s=1}}^{S}\!\sum_{i=1 }^{K^{(s)}}\!p_{si}\mathbb{E} \{ \mid{\bf g}_{si}^{\text {H}}{\bf h}^{s}_{jk}\! \mid ^{2}\}\!-\!p_{jk}\!\mid \mathbb{E} \{{\bf g}_{jk}^{\text {H}} {\bf h}^{j}_{jk}\}\mid ^{2}}_{\text{Interference power}}\!+\!\underbrace{\! \sigma ^{2} }_{\text{Noise power}}}.
	\end{equation}
	\end{small}
	The total sum achievable data rate for all the UEs in the $j$-th subarray can be denoted by

	\begin{equation}
		\text {SE}_{j} =\sum_{k=1}^{K^{(j)}} \mathbb{E}\left \{\log_{2}{(1+\text{SINR}_{jk} )}\right \}.
	\end{equation}
  The SE expression above is more general than the one without considering the interference between subarrays in \cite{9145378}. We notice that the key to design algorithms is to suppress the interference to obtain higher SE performance.
  
  \begin{figure}[t]
		\centering
		\includegraphics[width=2.8 in]{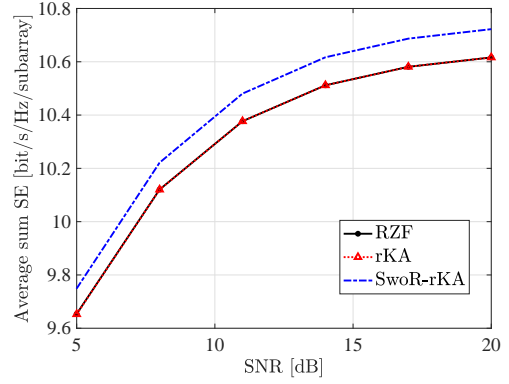}
		\caption{Average sum SE of the RZF algorithm, the rKA algorithm, and the SwoR-rKA algorithm against the SNR over perfect non-stationary channels in XL-MIMO systems with $M=256$, $K=16$, $S=4$.}
		\label{fig:sumrate1}
	\end{figure}
	\begin{figure}[t]
		\centering
		\includegraphics[width=2.8 in]{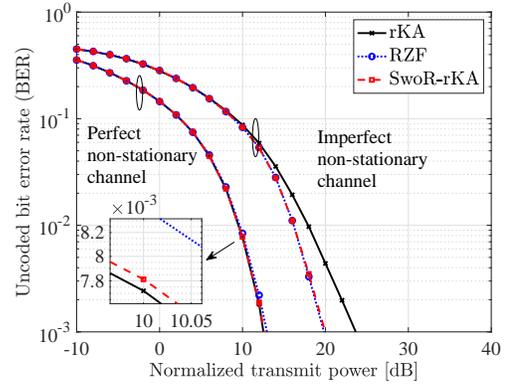}
		\caption{BER of the RZF algorithm, the rKA algorithm, and the SwoR-rKA algorithm against the normalized transmit power over perfect and imperfect non-stationary channels in XL-MIMO systems with $M=256$, $K=16$, $S=4$.}
		\label{fig:ber1}
	\end{figure}
  	\section{Simulation Results}\label{sec:simulation}
	
	In this section, we compare the performance between the traditional RZF algorithm, the proposed rKA algorithm, and the SwoR-rKA algorithm for the downlink precoding techniques in terms of sum-rate and computational complexity. Besides, we compare the convergence speed of the rKA algorithm and the SwoR-rKA algorithm. The simulation parameters are disposed in Table \ref{tab:Simulation parameters}.
	
	\begin{figure}[t]
		
		\centering
		\includegraphics[width=2.8 in]{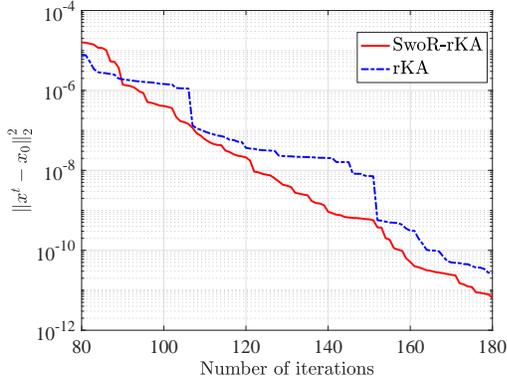}
		\captionsetup[figure]{justification=raggedright}
		\caption{NMSE of the rKA algorithm and SwoR-rKA algorithm with matrix inverse against the number of iterations in XL-MIMO systems with $M=256$, $K=16$.}
		\label{fig:norm2}
	\end{figure}

	\begin{figure}[htbp]
\centering
\subfigure[Number of complex multiplications against number of UEs.]{\includegraphics[height=3.5cm,width=4.35cm]{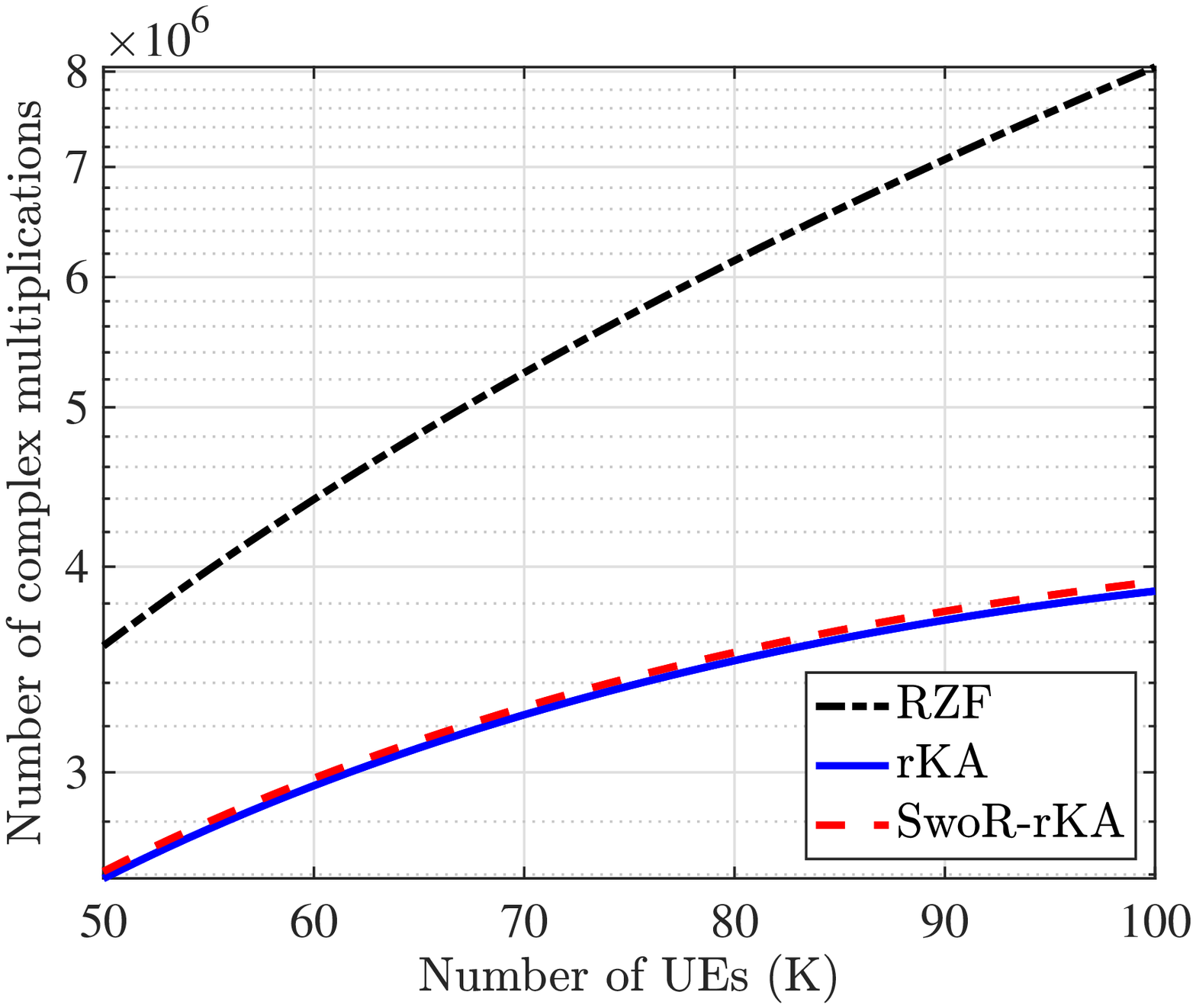}}
\subfigure[Number of complex multiplications against number of antennas.]{\includegraphics[height=3.5cm,width=4.35cm]{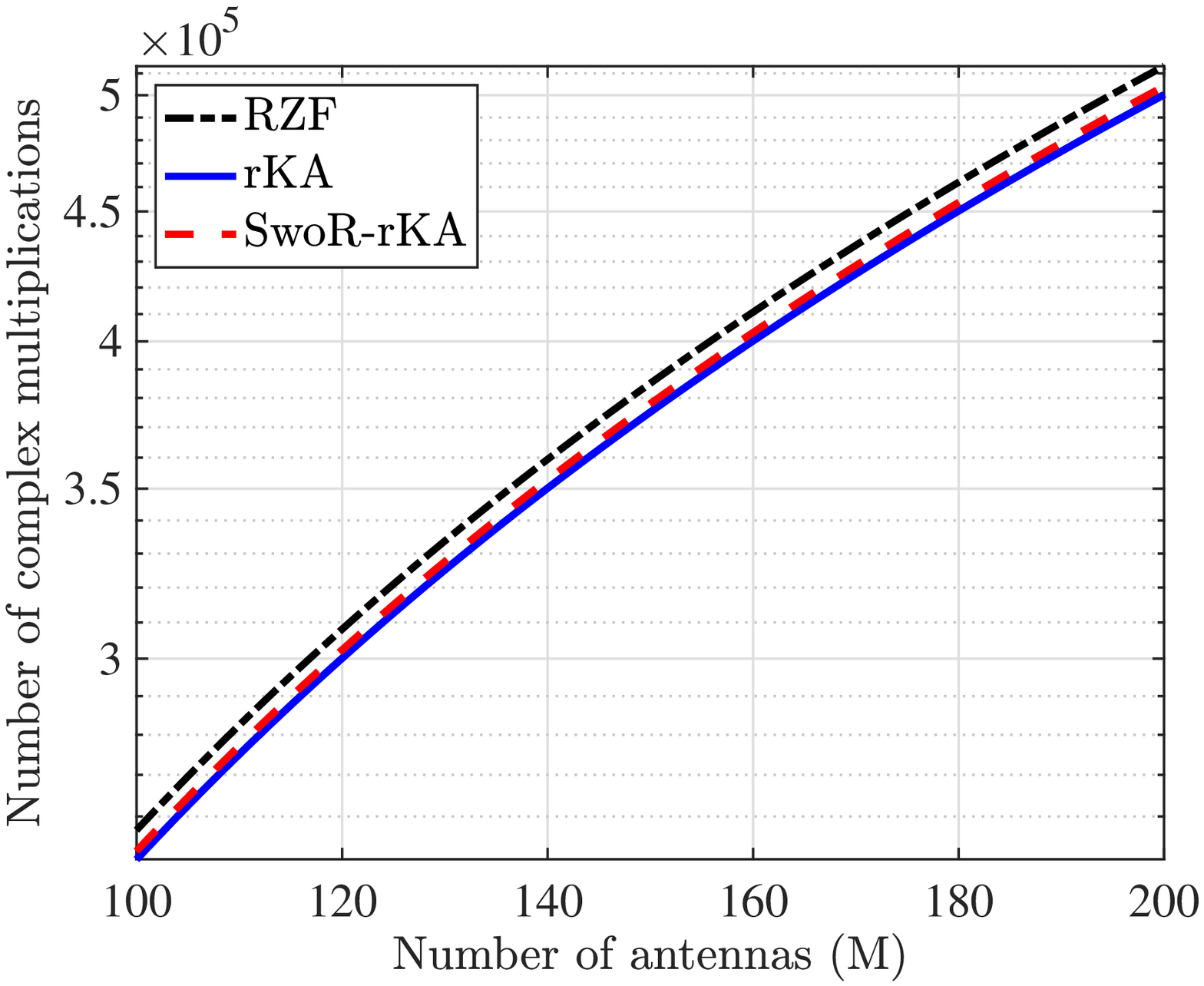}}
\caption{Complexity of the RZF algorithm, the rKA algorithm, and the SwoR-rKA in XL-MIMO systems.}
\label{complex}
\end{figure}	
	
	In Fig. \ref{fig:sumrate1}, we plot the sum rate with the RZF algorithm, the rKA algorithm, and the SwoR-rKA algorithm for non-stationary channels. It can be seen from Fig. \ref{fig:sumrate1} that the proposed SwoR-rKA algorithm achieves the best performance. Compared with the RZF precoding, the sum rate of the SwoR-rKA algorithm has about $0.9 \%$ improvement in non-stationary channels. Although they both obtain with almost the same SE, our algorithm has a lower complexity, as will be further investigated latter. Besides, the RZF algorithm and the rKA algorithm have almost the same SE performance.
 	
	Fig. \ref{fig:ber1} shows that the BER as a function of the SNR for the RZF algorithm, the rKA algorithm, and the SwoR-rKA algorithm in perfect and imperfect non-stationary channels  with $M=256, K=16$ and scalar parameter $\tau=0.3$ indicating the quality of the instantaneous CSI. When the SNR is $6$ dB in perfect channel conditions, we notice that the SwoR-rKA algorithm and the rKA algorithm have $2.2 \%$ improvement than the RZF precoding. Besides, the accuracy of the rKA algorithm and the SwoR-rKA algorithm is basically the same. Besides, the randomization method of SwoR-rKA algorithm makes it closer to the traditional RZF algorithm than rKA algorithm to solve the system of equations. It is obviously that the SwoR-rKA algorithm can effectively reduce the BER when the transmitter has an imperfect channel estimate.
	
	In Fig. \ref{fig:norm2}, we consider the normalized mean square error (NMSE) of the rKA algorithm and SwoR-rKA algorithm with matrix inverse against the number of iterations in XL-MIMO systems with $M=256$, $K=16$. When the number of iterations is $100$, the rKA algorithm and SwoR-rKA achieve $1.4 \times 10^{-6}$ NMSE and $3.7 \times 10^{-7}$ NMSE, respectively, representing an excellent result in computing the precoding vector. Besides, it can be seen that the SwoR-rKA algorithm has a faster convergence speed than the rKA algorithm.
	
	In Fig. \ref{complex}(a), we plot the complexity of the RZF algorithm, the rKA algorithm, and the SwoR-rKA algorithm against the number of UEs in XL-MIMO systems with $M=100$. With $M=100, K=100$, the complexity given by the rKA algorithm and the SwoR-rKA have about $51.3 \%$ reduction than the traditional RZF algorithm. Moreover, we can observe a trade-off between the performance and complexity of the algorithm: the SwoR-rKA algorithm has higher SE than the rKA algorithm, but the latter has lower complexity, which means more efficient execution. Besides, we notice that with the increase of the number of UEs, the gap between the RZF precoding and our proposed algorithms gradually increases. In Fig. \ref{complex}(b), we plot the complexity of the RZF algorithm, the rKA algorithm, and the SwoR-rKA algorithm against the number of antennas in the XL-MIMO system with $K=10$. The complexity given by the rKA algorithm and the SwoR-rKA are about $4.8 \%$ times lower than traditional RZF algorithm. Besides, the complexity of the rKA algorithm and the SwoR-rKA is almost identical.

%
%
%
%
%

	TABLE \ref{tab:complexity} summarizes the complexity of all the considered schemes, in two typical scenarios, i.e., the M-MIMO and the XL-MIMO. We investigate that in M-MIMO $M=64$, $K=8$ and in XL-MIMO $M=256$, $K=16$, and $S=4$. In contrast, the rKA and SwoR-rKA have lower algorithm complexity than the RZF algorithm, especially with an extremely large number of antennas. Moreover, although the SwoR-rKA algorithm has higher complexity than the rKA algorithm, it can provide higher SE. Also, we find that the advantages of our algorithm are more significant when the number of antennas and users reaches a certain scale.

	\begin{table}[t]
		\centering
		\fontsize{9}{12}\selectfont
		\caption{Computational Complexity for Precoding Methods based on Complex Operations.}\label{tab:complexity}
		\begin{tabular}{|m{0.9 cm}<{\centering}|m{4cm}<{\centering}|m{0.9cm}<{\centering}|m{0.9cm}<{\centering}|}
			\hline
		 \bf Scheme &  \bf Computational Complexity & \bf M-MIMO& \bf XL-MIMO \cr\hline
		 \hline
			 		
			RZF & $S[\frac{3(K^{(s)})^{2}M^{(s)}  }{2}+\frac{3K^{(s)}M^{(s)}  }{2}+\frac{(K^{(s)})^{3}-K^{(s)}   }{3}] $    & 7584 & 109888 \cr\hline
			rKA &$S[M^{(s)}T^{(s)}+M^{(s)}]$& 12864 & 51456 \cr\hline
			SwoR-rKA &$S[M^{(s)}T^{(s)}+2M^{(s)}K^{(s)}] $ & 13824 & 59392 \cr\hline
		\end{tabular}
	\end{table}

	\section{Conclusion}\label{sec:conclusion}
	In this work, we proposed the rKA and SwoR-rKA approaches to hugely reduce the computational complexity of downlink precoding in XL-MIMO systems. It is clear that changing the mode of randomization can effectively improve the convergence speed of the algorithm.
	Through simulations, we demonstrated that the proposed method can provide lower complexity and higher SE performance in XL-MIMO downlink systems. An interesting analogy can be made in this paper, our algorithms can effectively keep the BER performance when the transmitter has an imperfect channel estimate. In future work, we will address the influence of channel spatial non-stationarity on the energy efficiency and consider other near-field characteristics of XL-MIMO systems.
	
	\appendix
	\begin{appendices}
		\subsection{Convergence Analysis of the rKA Algorithm}\label{secA}
		We first consider a SLE $\bf Ax=b$, where ${\bf A}\in \mathbb{C}^{m \times n}$, ${\bf b}\in \mathbb{C}^{m}$.
		In each iteration, Algorithm \ref{algo:review} selects a row of $\bf A$ according to a given probability distribution ${\bf p}=(p_{1},\cdot \cdot \cdot ,p_{m}  )^{\text {T}} $, where $p_{i}\in [0,1] $, and $ {\textstyle \sum_{i=1}^{i=m}}p_{i}=1 $.
		\begin{equation}
			{\bf x}_{k+1}={\bf x}_{k}+\frac{{\bf b}_{r(i)}-\left \langle {\bf a}_{r(i)}, {\bf x}_{k}   \right \rangle  }{\left \| {\bf a}_{r(i)}  \right \| }{\bf a}_{r(i)},
		\end{equation}
		where $r_{(i)}$ is chosen from the set $1,2,...,m$ with probability distribution ${\bf p}$.
	
		We denote the column span of ${\bf A}^{\text {H}}$ as $\chi $, and define the normalized minimum gain of the matrix $\bf A$ along the subspace $\chi$ as  \cite{boroujerdi2018low}:
		\begin{equation}
			k_{\chi }({\bf A} )=\min_{x\in \chi ,{\bf x}\ne 0 } \frac{\left \| {\bf A}{\bf x}  \right \|^{2} } {\left \| {\bf A} \right \|_{F}^{2} \left \| {\bf x} \right \|^{2}}.
		\end{equation}
		As such, we obtain the convergence rate of this strategy
		\begin{equation}
			\mathbb{E}\left \{ \left \| {\bf x}^{t}-{\bf x}^{\star }   \right \|^{2}   \right \}\le (1-k_{\chi }({\bf A} ))^{t}  \left \| {\bf x}^{0}-{\bf x}^{\star }   \right \|^{2},
		\end{equation}
		where ${\bf x}^{t}$ is the solution of the $t$-th iteration of the SLE $\bf Ax=b$ and ${\bf x}^{\star}$ is the actual solution.
		From (32), it is seen that the convergence speed of our algorithms depends on the parameter $k_{\chi }({\bf A} )$. Moreover, a larger $k_{\chi }({\bf A} )$ guarantees a faster convergence. Note that (19) and (23) represent two different randomization schemes, respectively. The latter tends to select the columns with larger values and it is found through simulation that its $k_{\chi }({\bf A}) $ is larger, i.e., its convergence speed is faster.

	\end{appendices}
	
	\bibliographystyle{IEEEtran}
	\bibliography{IEEEabrv,ref}

\end{document}